\documentclass[12pt,a4wide]{article}
\usepackage{amsmath,amsthm,epsfig,graphicx}
\usepackage{amsmath,amsthm,amssymb}
\usepackage{amsfonts}
\usepackage{graphicx}
\usepackage{latexsym}
\usepackage[mathcal]{eucal}
\usepackage{epsfig}
\usepackage{verbatim,url}
\renewcommand{\lambda}{\ell}

\newcommand{\Qx}{ \mathbb{Q} }
\newcommand{\Ex}{ \mathbb{E} }

\newcommand{\rec}{\mbox{R{\tiny EC}}}

\newcommand{\lgd}{\mbox{L{\tiny GD}}}

\newcommand{\cds}{\mbox{CDS}}

\newcommand{\pcds}{\Pi\mbox{\tiny RCDS}}

\numberwithin{equation}{section}

\title{\vspace{-2cm} {\normalsize  An extended and updated version of this paper with the title} \\
 {\normalsize \bf Liquidity modeling for Credit Default Swaps: an overview} \\ {\normalsize will appear in: Bielecki, T., Brigo, D., and Patras, F. (Editors)}\\ {\normalsize ``Credit Risk Frontiers. The subprime crisis, Pricing and Hedging, CVA,
MBS, Ratings and Liquidity", Bloomberg Press.}\\ --- \\
{\bf \Large Credit Default Swaps Liquidity modeling:\\ A survey\thanks{We are grateful to Rutang Thanawalla, Greg Gupton, Wei Liu, Ahmet Kocagil and Alexander Reyngold, who worked with us on the methodology of the Fitch Solutions liquidity project, contributing to our insight in this challenging field. This work expresses the opinion of its authors and is in no way representing the opinion of the institutions the authors work for}}}
\author{Damiano Brigo \  Mirela Predescu \  Agostino Capponi\thanks{Brigo is with the Dept of Mathematics, Imperial College, London, {\tt d.brigo@imperial.ac.uk}; Predescu is with Lloyds TSB, Portfolio Policy, Analytics and Optimization Group, {\tt mirela.predescu@lloydsbanking.com};  Capponi holds a PhD from Caltech, {\tt acapponi@caltech.edu }. First Version: February 22, 2010.
}
}

\date{}

\begin{document}

\maketitle

\thispagestyle{empty}

\newpage

\begin{abstract}
We review different theoretical and empirical approaches for measuring the impact of liquidity on CDS prices. We start by reduced form models incorporating liquidity as an additional discount rate. We review Chen, Fabozzi and Sverdlove (2008) and Buhler and Trapp (2006, 2008), adopting different assumptions on how liquidity rates enter the CDS premium rate formula, about the dynamics of liquidity rate processes and about the credit-liquidity correlation. Buhler and Trapp (2008) provides the most general and realistic framework, incorporating correlation between liquidity and credit, liquidity spillover effects between bonds and CDS contracts and asymmetric liquidity effects on the Bid and Ask CDS premium rates.
We then discuss the Bongaerts, De Jong and Driessen (2009) study which derives an equilibrium asset pricing model incorporating liquidity effects. Findings include that both expected illiquidity and liquidity risk have a statistically significant impact on expected CDS returns, but only compensation for expected illiquidity is economically significant with higher expected liquidity being associated with higher expected returns for the protection sellers. This finding is contrary to Chen, Fabozzi and Sverdlove (2008) that found protection buyers to earn the liquidity premium instead.
We finalize our review with a discussion of Predescu et al (2009), which analyzes also data in-crisis. This is a statistical model that associates an ordinal liquidity score with each CDS reference entity and allows one to compare liquidity of over 2400 reference entities. This study points out that credit and illiquidity are correlated, with a smile pattern.
All these studies highlight that CDS premium rates are not pure measures of credit risk. CDS liquidity varies cross-sectionally and over time.  CDS expected liquidity and liquidity risk premia are priced in CDS expected returns and spreads.  Further research is needed to measure liquidity premium at CDS contract level and to disentangle liquidity from credit effectively.

\end{abstract}

{\bf AMS Classification Codes}: 60H10, 60J60, 91B70;


\bigskip


\newpage

\tableofcontents

\newpage

\pagestyle{myheadings}
\markboth{}{{\footnotesize  D. Brigo, M. Predescu, A. Capponi. CDS Liquidity Modeling survey}}

\section{Introduction}

Liquidity is a notion that has gained increasing attention following the credit crisis that started in 2007 (``the crisis" in the following). As a matter of fact, this has been a liquidity crisis besides a credit crisis. For many market players, problems have been aggravated by the lack of reserves when in need to maintain positions in order to avoid closing deals with large negative mark to markets. This lack of reserves forced fire sales at the worst possible moments and started a domino effect leading to the collapse of financial institutions.

\subsection{Funding, Trading and Systemic Liquidity}
Szego (2009) illustrates, among other factors, a negative loop involving illiquidity as fueling the crisis development. We can consider for example the following schematization:

\begin{itemize}
\item[1.] (Further) liquidity reduction on asset trade;
\item[2.] (Further) price contraction due to liquidity decline;
\item[3.] (Further) decline of value of bank assets portfolio;
\item[4.] Difficulty in refinancing, difficulty in borrowing, forced to (further) sale of assets;
\item[5.] Assets left? If so, go back to 1. If not:
\item[6.] Impossibility of refinancing;
\item[7.] Bankruptcy.
\end{itemize}

This sketchy and admittedly simplified representation highlights the three types of liquidity that generally market participants care about. One is the market/trading liquidity generally defined as the ability to trade quickly at a low cost (O'Hara (1995)). This generally means low transaction costs coming from bid-ask spreads, low price impact of trading large volumes and considerable market depth.  This notion of market liquidity can be applied to different asset classes (equities, bonds, interest rate products, FX products, credit derivatives etc.) and to the overall financial markets. In addition to trading liquidity, banks and financial institutions also closely monitor the funding liquidity, which is the ease with which liabilities can be funded through different financing sources.

Market and funding liquidity are related since timely funding of liabilities relies on the market liquidity risk of its assets, given that a bank may need to sell some of its assets to match its liability-side obligations at certain points in time. The recent crisis prompted regulators and central banks to look very closely at both types of liquidity and to propose new guidelines for liquidity risk management (see BIS(2008), FSA(2009)).

A third kind of liquidity that is however implicit in the above schematization, is the systemic liquidity risk associated to a global financial crisis, characterized by a generalized difficulty in borrowing.

%
%


As with other types of risks, liquidity needs to be analyzed from both a pricing perspective and a risk management one.

\subsection{Liquidity as a pricing component}

In the pricing space, Amihud, Mendelson, and Pedersen (2005) provide a thorough survey of theoretical and empirical papers that analyze the impact of liquidity on asset prices for traditional securities such as stocks and bonds. Other papers (Cetin, Jarrow, Protter, and Warachka (2005),  Garleanu, Pedersen and Poteshman (2006)) investigated the impact of liquidity on option prices. More generally Cetin, Jarrow and Protter (2004) extends the classical arbitrage pricing theory to include liquidity risk by considering an economy with a stochastic supply curve where the price of a security is a function of the trade size. This leads to a new definition of self-financing trading strategies and to additional restrictions on
hedging strategies, all of which have important consequences in valuation. Their paper also reports a good summary of earlier literature on transaction costs and trade restrictions, to which we refer the interested reader.

\begin{figure}
\begin{center}
\includegraphics[scale=0.5]{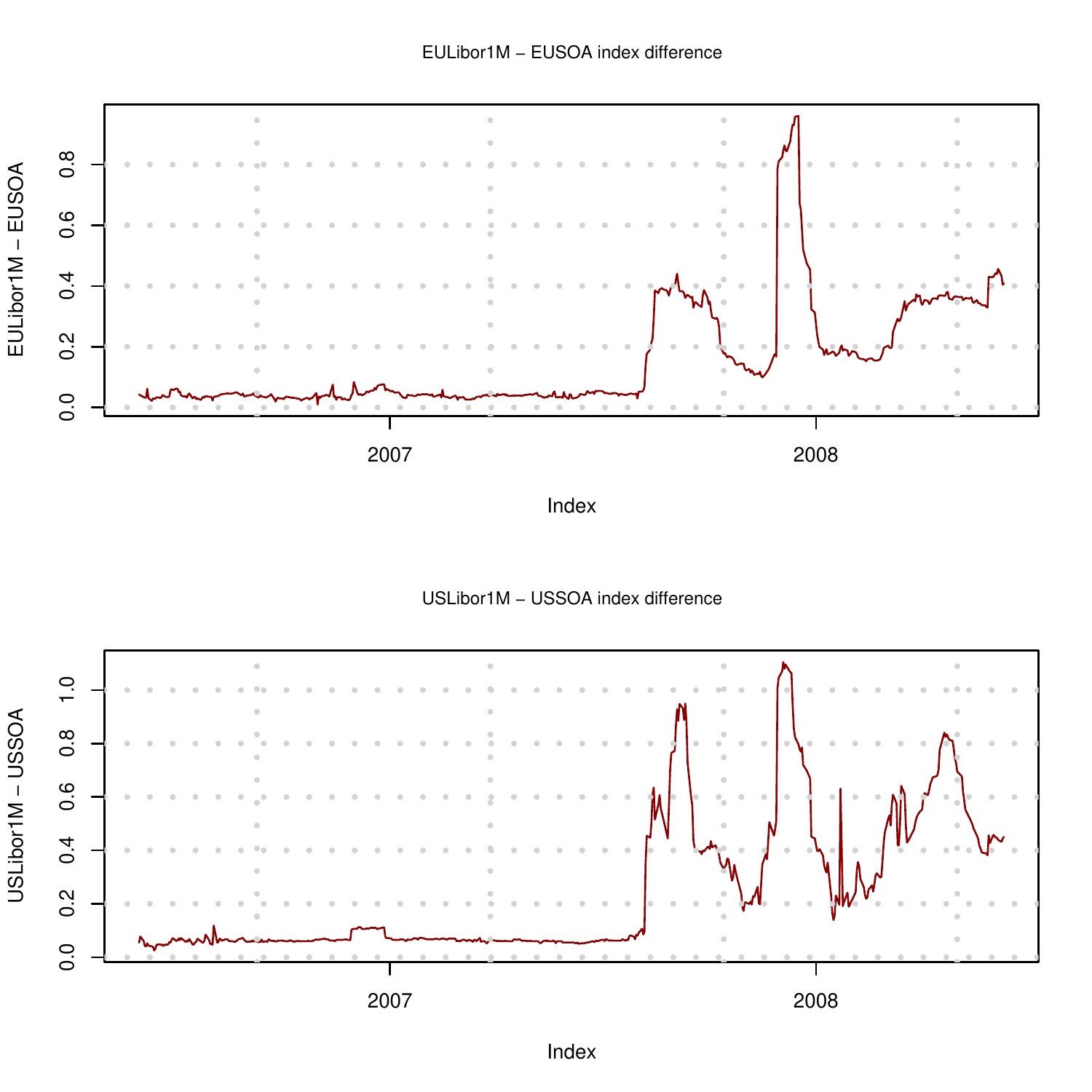}
\caption{Difference between 1 month EU (US) Libor and EU (US) overnight index swap}
\label{fig:liboreonia}
\end{center}
\end{figure}

Morini (2009) analyzes the liquidity and credit impact on interest rate modeling, building a framework that consistently accounts for the divergence between market forward rate agreements (FRA) rates and the LIBOR replicated FRA rates. He also accounts for the divergence between overnight indexed swap rates (EONIA) and LIBOR rates. The difference between the two rates can only be
attributed to liquidity or counterparty risk, the latter being almost zero in EONIA due to the very short (practically daily) times between payments. For illustration
purposes, we report in Fig.~\ref{fig:liboreonia} the differences between EONIA and LIBOR rates for Europe and the analogous difference for the United States.
It is clear from the graphs in Fig.~\ref{fig:liboreonia} that starting from the end of 2007 and till mid 2008, there is a noticeable
increase of the difference between 1 month Libor and overnight index swap rate, which is instead very small in the beginning of
the observation period. This is not surprising as the end of 2007 corresponds to the start of the subprime mortgage crisis, which
then exacerbated and became the credit crunch crisis for the all period of 2008.

The analysis done by Morini (2009) makes use of Basis Swaps between LIBOR with different tenors, and takes into account collateralization. Morini is able to reconcile the divergence in rates even under simplifying assumptions. His analysis however implicitly views liquidity as a consequence of credit rather than as an independent variable, although he does not exclude the possibility that liquidity may have a more independent component.

\subsection{Liquidity in Risk Measurement}

Several studies (Jarrow and Subramanian(1997), Bangia et al. (1999), Angelidis and Benos (2005), Jarrow and Protter(2005), Stange and Kaserer(2008), Earnst, Stange and Kaserer(2009) among few others) propose different methods of accounting for liquidity risk in computing risk measures. Bangia et al. (1999) classify market liquidity risk in two categories: (a) the exogenous illiquidity which depends on general market conditions, is common to all market players and is unaffacted by the actions of any one participant and (b) the endogenous illiquidity that is specific to one's position in the market, varies across different market players and is mainly related to the impact of the trade size on the bid-ask spread.  Bangia et al. (1999) and Earnst et al. (2009) only consider the exogenous illiquidity risk and propose a liquidity adjusted VaR measure built using the distribution of the bid-ask spreads. The other mentioned studies model and account for endogenous risk in the calculation of liquidity adjusted risk measures.

In the context of the coherent risk measures literature, the general axioms  a liquidity measure should satisfy are discussed in Acerbi and Scandolo (2008). They propose a formalism for Liquidity Risk which is compatible with the axioms of coherency of the earlier risk measures literature. They emphasize the important but easily overlooked difference between coherent risk measures defined on portfolio values and coherent risk measures defined on the vector space of portfolios. The key observation is that in presence of liquidity risk the value function on the space of portfolios is not necessarily linear. From this starting point a theory is developed, introducing a nonlinear value function depending on a notion of liquidity policy based on a general description of the microstructure of illiquid markets and the impact that this microstructure has when marking to market a portfolio.

\subsection{This survey: Liquidity in CDS pricing}

In this paper we focus on liquidity modeling in the valuation space, to which we go fully back now, and more specifically, in the context of credit derivatives instruments, on the impact of liquidity on credit default swaps (CDS) premium rates. CDS represent the most liquid credit instruments and are highly standardized. The basic idea to include liquidity as a spread, leading to a liquidity stochastic discount factor, follows the approach adopted for example by Chen, Cheng and Wu (2005)\footnote{this is reviewed in Brigo, Predescu and Capponi (2010) but not here}, Buhler and Trapp (2006) and (2008) (BT06 and BT08) and Chen, Fabozzi and Sverdlove (2008) (CFS), among others.
All approaches but BT08 are unrealistic in that they assume the liquidity rate to be independent of the hazard rate associated to defaults of the relevant CDS entities. BT08 and Predescu et al (2009) show, although in different contexts, that liquidity and credit are correlated. We discuss their results.

We will then analyze a different approach, by Bongaerts, De Jong and Driessen (2009) (BDD) who use Capital Asset Pricing Model (CAPM) like arguments to deduce liquidity from CDS data. None of these works uses data in-crisis, i.e. after June 2007.
One exception is Predescu et al (2009) (PTGLKR), where liquidity scores for CDS data are produced starting from contributors BID ASK or MID CDS quotes across time. This is an ordinal liquidity measure,
as opposed to a more attractive cardinal one, but it represents - to the best of our knowledge - the only work dealing with CDS liquidity using also crisis data. After the ordinal model by Predescu et al (2009), we go back to cardinal models and briefly hint at Tang and Yan (2007), that also includes bid ask information among other variables chosen as liquidity measures: volatility to volume, number of contracts outstanding, trade to quote ratio and volume. Tang and Yan is reviewed more in detail in Brigo, Predescu and Capponi (2010).

We then conclude the paper by comparing the often contradictory findings of the above works, pointing out remaining problems in liquidity estimation and pricing in the credit market.


\section{Liquidity as a spread in Reduced Form Models}
The basic idea in this approach is to include liquidity as a (possibly stochastic)  spread, leading to a liquidity (possibly stochastic) discount factor.

In order to be able to work on liquidity for CDS we need to introduce the CDS contract and its mechanics. To this end we follow Brigo and Mercurio (2006), Chapter 21.

\subsection{Credit Default Swaps}

The running CDS contract is defined as follows.  A CDS contract ensures protection against default.  Two parties
``A" (Protection buyer) and ``B" (Protection seller) agree on the following.

If a third party ``C" (Reference Credit) defaults at time $\tau=\tau_C$, with $T_a < \tau <T_b$, ``B" pays to ``A" a certain  amount $\lgd$ (Loss Given Default of the
reference credit ``C"). In turn, ``A" pays to "B" a premium rate $S$ at
times $T_{a+1},\ldots,T_b$ or until default. Set $\alpha_i = T_i -
T_{i-1}$ and $T_0=0$. We can summarize the above structure as%

\bigskip

\[ \fbox{$\begin{array}{c} \mbox{Protection} \\ \mbox{Seller B} \end{array}$} \begin{array}{ccc}  \rightarrow &
\mbox{ protection } \lgd \mbox{ at default $\tau_C$ if $T_a<
\tau_C \le T_b$}
& \rightarrow  \\
%
 \leftarrow  & \mbox{ rate } S \mbox{ at }
T_{a+1},\ldots,T_b \mbox{ or until default } \tau_C & \leftarrow
\end{array} \fbox{$\begin{array}{c} \mbox{Protection} \\ \mbox{Buyer A } \end{array}$} \]

\bigskip

(protection leg and premium leg respectively). The amount $\lgd$ is a {\em
protection} for ``A" in case ``C" defaults. Typically $\lgd=$
notional, or ``notional - recovery" $=1
-\rec$.

Formally, we may write the CDS discounted value at time $t$ as seen from ``B" as
\begin{eqnarray}\label{eq:discountedpayoffcds}
\pcds_{a,b}(t) :=  D(t,\tau) (\tau-T_{\beta(\tau)-1}) S
\mathbf{1}_{\{T_a < \tau < T_b \} }\\ \nonumber
 +   \sum_{i=a+1}^b D(t,T_i) \alpha_i S \mathbf{1}_{\{\tau \ge T_i\}
 }- \mathbf{1}_{\{T_a < \tau \le T_b \} }D(t,\tau) \ \lgd
\end{eqnarray}
where $t\in [T_{\beta(t)-1},T_{\beta(t)})$, i.e. $T_{\beta(t)}$ is
the first date among the $T_i$'s that follows $t$ and $D(t,T)$ is the risk free discount factor at time $t$ for maturity $T$.

A note on terminology: In the market $S$ is usually called ``CDS spread". However, we will use ``spread" both to denote the difference between the Ask and Bid quotes of a security and to indicate instantaneous rates on top of the default-free instantaneous rate $r$. To avoid confusion we will refer to $S$ as to the CDS premium rate rather than CDS spread.

Usually, at inception time (say $0$) the amount $S$ is set at a
value  $S_{a,b}(0)$ that makes the contract fair, i.e. such that
the present value of the two exchanged flows is zero. This is how
the market quotes running CDS's: CDS are quoted via their fair $S$'s (Bid
and Ask).

Recently, there has been some interest in ``upfront CDS" contracts with a fixed premium rate in the premium leg.\footnote{See for example Beumee et al (2009) for a discussion on the upfront features and on the running-upfront conversion.}
In these contracts the premium rate $S$ is fixed to some pre-assigned canonical value $\bar{S}$, typically 100 or 500 basis points (bps, 1bp$= 10^{-4}$), and the remaining part of the protection is paid upfront by the party that is buying
protection. In other terms, instead of exchanging a possible
protection payment for some coupons that put the contract in equilibrium at inception, one exchanges it with a fixed coupon and compensates for the difference with an
upfront payment.

%
%

We denote by $\cds(t,[T_{a+1},\ldots,T_b],T_a,T_b, S, \lgd)$ the
price at time $t$ of the above standard running CDS~(\ref{eq:discountedpayoffcds}).
At times some
terms are omitted, such as for example the list of payment dates
$[T_{a+1},\ldots,T_b]$, and to shorten notation further we may
write $\cds_{a,b}(t, S, \lgd)$.

The pricing formulas for these payoffs depend on the assumptions
on interest-rate dynamics and on the default time $\tau$. If
$\tau$ is assumed to be independent of interest rates, then model
independent valuation formulas for CDS's involving directly
default (or survival) probabilities and default free zero coupon
bonds are available.

Proper valuation of CDS should also take into account counterparty risk, see for example Brigo and Chourdakis (2009, unilateral case) and Brigo and Capponi (2008, bilateral case). Here however we focus on works that did not consider counterparty risk in the CDS valuation.

In general, whichever the model,  we can compute the CDS price
according to risk-neutral valuation:
\begin{eqnarray}\label{CDSriskneutralprice}
\cds_{a,b}(t, S, \lgd) = \Bbb{E}_t \{\pcds_{a,b}(t) \} .
\end{eqnarray}
where $\Ex_t$ is the risk neutral expectation conditional on the market information at time $t$.

If we define the fair premium of the CDS at a given time $t$ as the value of the premium rate $S =  S_{a,b}(t)$ such that
$\cds_{a,b}(t, S_{a,b}(t), \lgd) = 0$, i.e. such that the two legs of the CDS have the same value, we can write,
on $\{\tau > t\}$,

\begin{equation}\label{eq:cdsfairspread}
S_{a,b}(t) = \frac{\Ex_t[\lgd\ \mathbf{1}_{\{T_a < \tau \le T_b \} }D(t,\tau)]}{\Ex_t[D(t,\tau) (\tau-T_{\beta(\tau)-1})
\mathbf{1}_{\{T_a < \tau < T_b \} }
 +   \sum_{i=a+1}^b D(t,T_i) \alpha_i  \mathbf{1}_{\{\tau \ge T_i\} }]}
\end{equation}

If we assume independence between rates and the default time,
or more in particular deterministic interest rates,
then the default time $\tau$ and interest rate quantities $r, D(s,t),...$ are independent.
It follows that the (receiver) CDS valuation, for a CDS selling protection at time $0$
for defaults between times $T_a$ and $T_b$ in exchange of a periodic premium rate $S$ becomes
(PL = Premium Leg, DL = Default Leg or Protection Leg)
%
\begin{eqnarray}\label{eqn:credit:modindcdstot}
\hspace{-1cm} \cds_{a,b}(0, S, \lgd;\Qx(\tau > \cdot))  = \mbox{PL}_{a,b}(0,S;\Qx(\tau > \cdot)) -
  \mbox{DL}_{a,b}(0,\lgd;\Qx(\tau > \cdot))\hspace{1cm} \\ \label{eqn:credit:modindcdspl}
\mbox{PL}_{a,b}(0,S)=     S \left[ -\int_{T_a}^{T_b} P(0,t) (t-T_{\beta(t)-1})
d_t \Qx(\tau \ge t) +   \sum_{i=a+1}^b P(0,T_i)  \alpha_i   \Qx(\tau \ge
 T_i) \right] \\ \label{eqn:credit:modindcdsdl}
\mbox{DL}_{a,b}(0,\lgd) =  - \lgd \left[
\int_{T_a}^{T_b} P(0,t)\  d_t \Qx(\tau \ge t)\right],
\end{eqnarray}
where $P(t,u) = \Ex_t [D(t,u)]$ is the price of a default risk free zero coupon bond. In case rates are deterministic we have $D(t,u) = P(t,u)$ for all $t,u$.
The above CDS formula is model independent.
In particular, for a spot CDS with $T_a=0$ at $t=0$ we have that the fair premium rate formula
(\ref{eq:cdsfairspread}) becomes
\begin{equation}\label{eq:cdsfairspread0}
S_{0,b}(0) = \frac{- \lgd
\int_{0}^{T_b} P(0,t)\  d_t \Qx(\tau \ge t) }{-\int_{0}^{T_b} P(0,t) (t-T_{\beta(t)-1})
d_t \Qx(\tau \ge t) +   \sum_{i=1}^b P(0,T_i)  \alpha_i   \Qx(\tau \ge
 T_i) }
\end{equation}

This means that if we strip survival probabilities from
CDS in a model independent way at time 0, to calibrate a chosen model for $\tau$ to the market CDS quotes
we just need to make sure that the survival probabilities we strip
from CDS are correctly reproduced by the chosen $\tau$ model, whichever it is.

Equations~(\ref{eqn:credit:modindcdspl}, \ref{eqn:credit:modindcdsdl}) are no longer valid in general if we remove the independence  between $\tau$ and interest rates. This complicates matters considerably and this is the reason why most of the works on liquidity risk with CDS tend to assume independence between default-free interest rates and default.

Most of the approaches to liquidity we discuss in this paper are based on the intensity approach (or reduced form approach) to modeling $\tau$. In this approach, default is not induced by basic market observables and/or economic
fundamentals, but has an exogenous component that is independent of all the default free market information. Monitoring the default free market (interest rates, exchange rates, etc) does not give complete information on the default process, and there is no economic rationale behind default. This family of models is particularly suited to model credit spreads and in its basic formulation is easy to calibrate to Credit Default Swap (CDS) or corporate bond data.

\subsection{Intensity models for CDS}

We need now to introduce intensity models. First we need a few definitions from probability theory.
We place ourselves in a probability space $(\Omega,\mathcal{G},\mathcal{G}_t,\mathbb{Q})$.
The filtration $(\mathcal{G}_t)_t$ models the flow of information
of the whole market, including credit and defaults. $\Qx$ is the risk neutral
measure. This space is endowed also with a right-continuous and
complete sub-filtration $\mathcal{F}_t$ representing all the
observable market quantities but the default events (hence
$\mathcal{F}_t\subseteq\mathcal{G}_t:=\mathcal{F}_t\vee\mathcal{H}_t$
where $\mathcal{H}_t=\sigma(\{\tau \leq u\}:u\leq t)$ is the
right-continuous filtration generated by the default event).

We set $\mathbb{E}_t(\cdot):=\mathbb{E}(\cdot|\mathcal{G}_t)$, the
risk neutral expectation leading to prices.

Intensity models are based on the assumption that the default time $\tau$ is the first
jump of a Cox process with intensity $h_t$ (see for example Bielecki and Rutkowski (2001)
for more details, or Brigo and Mercurio (2006)).  This in particular implies
\[\Qx\{\tau \in [t,t+dt) |\tau \ge t, {\cal F}_t\} = h_t \ dt,\]
where the stochastic process $h$ is the intensity.
This reads, if ``$t$=now":\\
{\em ``probability that reference entity defaults in (small) $``dt"$  years
given that it has not defaulted so far and given the
default-free-market information so far is $h_t\ dt$."}

Intensity is usually assumed to be at least a ${\cal F}_t$-adapted and right
continuous (and thus progressive) process and is denoted by
$h_t$, and the {\em cumulated intensity} or {\em hazard
process} is the stochastic process  $T \mapsto H(T) = \int_0^T h_t
dt$.
We assume $h_t >0$. We recall that the requirement to be
``${\cal F}_t$-adapted"  means that given ${\cal F}_t$, i.e. the
default-free market information up to time $t$, we know $h$
from $0$ to $t$. This intuitively says that the randomness we
allow into the intensity is induced by the default free market. In
a {\em Cox process} with stochastic intensity $h$,
conditional on ${\cal F}_t$  (or just on ${\cal F}_t^h=\sigma(\{h_s : s \le t\})$, i.e. just on the
paths of $h$), we have a Poisson process structure with intensity $h_t$.
In particular, we have that the first jump time of
the process, transformed through its cumulated intensity, is an
exponential random variable independent of ${\cal F}_t$:
\[H(\tau) = \xi\]
with $\xi$ standard (unit-mean) exponential random variable independent of ${\cal F}_t$. Then
we have that default can be defined as  \[\tau :=
H^{-1}(\xi), \]
that provides us with suggestions on how to simulate the
default time in this setting.

Notice that in this setting not only $\xi$ is random, but $h$
itself is stochastic. This is why Cox processes are at times
called ``doubly stochastic Poisson processes".

For the survival probability in intensity models we have
\[\Bbb{Q}\{\tau \ge t \} = \Bbb{Q}\{H(\tau) \ge H(t) \} = \Bbb{Q}\bigg{\{}\xi \ge \int_0^t h(u)du
\bigg{\}}\]\[= \Ex\bigg{[} \Bbb{Q}\bigg{\{}\xi \ge \int_0^t
h(u)du \bigg{|} {\cal F}_t^h \bigg{\}}\bigg{]}= \Ex
\bigg{[} e^{-\int_0^t h(u)du} \bigg{]}\] which is completely
analogous to the bond price formula in a short interest-rate model with
interest rate $h$. This is the reason why the intensity can be seen also as an instantaneous credit spread.
This is further illustrated by computing the price of a zero coupon defaultable bond in the intensity setting.
The price at time $t$ of a zero coupon defaultable bond with maturity $T$ and zero recovery turns out to be
\begin{equation}\label{eq:DEFZCB} \Bbb{E}\{ D(t,T) 1_{\{ \tau
> T \} }|{\cal G}_t \} = 1_{\{ \tau
> t \} }  \Ex
\bigg{[} e^{-\int_t^T (r_u + h_u ) \ du }|{\cal F}_t \bigg{]}
\end{equation}
where $r_t$ is the instantaneous (possibly stochastic) default-free interest rate at time $t$,
so that $D(t,u) = \exp(-\int_t^u r_s ds)$.

Cox processes allow to borrow the stochastic interest-rates
technology and paradigms into default modeling, although
$\xi$ remains independent of all default free market quantities (of
${\cal F}$, of $h$, of $r$...) and represents an external
source of randomness that makes reduced form models incomplete.

In this setting, the time varying nature of $h$ may account for the term
structure of credit spreads, while the stochasticity of $h$
can be used to introduce credit spread volatility. For example, in
a diffusion setting, one has

\[ d h_t = b(t,h_t) dt + \sigma(t, h_t)\ dW_t  \]

for suitable choices of the drift $b$ and diffusion coefficient $\sigma$ functions.
$b$ and $\sigma$ can be chosen so as to reproduce desired levels of credit spreads and credit spreads volatilities.

Intensity models have been introduced for example in the seminal work of Duffie and Singleton (1999).
For tractable models and their calibration to the CDS term structure also in relation with CDS option pricing see for example Brigo and Alfonsi (2005) and Brigo and El-Bachir (2008).

\subsection{Intensity models for liquidity}

In most of the models we will review in this paper liquidity enters the picture as a further spread, or intensity.
The idea is  starting from a formula like~(\ref{eq:DEFZCB}) and adding a new rate term
$\lambda_t$ to discounting.


In other terms, the price of a Zero Coupon Bond changes from (\ref{eq:DEFZCB}),
where no liquidity component is included, to
\begin{equation}\label{eq:DEFZCBbis} \Bbb{E}\{ D(t,T) L(t,T) 1_{\{ \tau
> T \} }|{\cal G}_t \} = 1_{\{ \tau
> t \} }  \Ex
\bigg{[} e^{-\int_t^T (r_u + h_u + \lambda_u ) \ du }|{\cal F}_t \bigg{]}
\end{equation}
where $L(t,u) = \exp(-\int_t^u \lambda_s ds)$ is a stochastic discount factor component due to illiquidity,
and $\lambda$ is the instantaneous rate associated to this discount.

We can get a feeling for this use of the spread with the following example.
As we have seen before, for running CDS's, the market quotes at a point in time the value of $S=S_{0,b}$ such that

\[ \cds_{0,b}(0,S_{0,b},\lgd) = 0\]

This quote may come with a Bid and an Ask quote. Let us call them $S^{Bid}$ and $S^{ask}$.
The (positive) difference $S^{ask}-S^{Bid}$ is called CDS BID-ASK spread, or sometimes more colloquially just ``CDS BID-ASK".

It is known that if we assume $h_t$ to be deterministic and constant in time,
and we take the premium leg of the CDS paying continuously instead of quarterly, the equilibrium
premium rate $S$ balancing the two legs at a given time is linked to $h$ by the formula
\begin{equation}\label{eq:flathazrate} h = \frac{S_{0,b}}{\lgd} \Rightarrow S_{0,b}  = h \lgd
\end{equation}
(see for example Brigo and Mercurio (2006)).
Now, since $S$ has a bid and an ask quote, we can apply this formula to both.
We obtain
\[ h^{ask} = S^{ask}_{0,b}/ \lgd , \ \ h^{bid} = S^{bid}_{0,b}/ \lgd . \]
One possible very rough first approach is to define the liquidity spread $\lambda$
(in this framework deterministic and constant)
as
\[ \lambda := \frac{h^{ask} -  h^{bid}}{2} . \]
If we do that, and we define
\[ h^{mid} := \frac{h^{bid}+h^{ask}}{2} =  \frac{S^{bid}_{0,b}+S^{ask}_{0,b}}{2 \lgd}    \]
we notice that by definition then
\[ h^{bid} = h_{mid} - \lambda, \ \ \ h^{ask} = h_{mid} + \lambda, \]
and using again formula (\ref{eq:flathazrate}),
\[ S^{mid}_{0,b} = {h^{mid}}{\lgd} = \frac{S^{bid}_{0,b}+S^{ask}_{0,b}}{2}  \]
so that we have consistency with a meaningful definition of quoted MID $S$.

According to this simple calculation, the true credit spread of a name is the MID, whereas the
range between BID and ASK is due to illiquidity. For a perfectly liquid name $\lambda = 0$ because the BID ASK spread is zero.
For a very illiquid name, with a large bid ask spread, $\lambda$ will be large.
So strictly speaking $\lambda$ would be an {\em illiquidity spread}, although it is called by abuse of language
``liquidity spread". In this paper whenever we mention ``liquidity spread" we mean an illiquidity spread. This means that if the spread increases liquidity decreases (illiquidity increases).

The above framework sees the MID CDS premium rate to be centered in the middle of the bid-ask range.
However, it does not follow that also the NPV's of the CDS position has the MID centered between BID and ASK.
This is illustrated in the above simplified framework with continuous payments in the premium
leg, flat hazard rates $h_t=h$ and constant deterministic interest rates $r$.
In that case, the CDS default leg NPV corresponding to bid and ask
becomes
\[ DL_{0,b}^{bid} =  \lgd\ h^{bid}\ \frac{(1 - \exp(-(r+h^{bid})T_b))}{r+h^{bid}} , \ \ \ DL_{0,b}^{ask} = \lgd\
 h^{ask}\  \frac{(1 - \exp(-(r+h^{ask})T_b))}{{r+h^{ask}} }  \]
and we can see that
\[  \lgd\ h^{mid}\ \frac{(1 - \exp(-(r+h^{mid})T_b))}{{r+h^{mid}}}  \neq \frac{DL_{0,b}^{bid} + DL_{0,b}^{ask}}{2}
\]
If we approximate the exponential at the first order we get back symmetry, but otherwise we do not observe that the NPV
corresponding to the CDS MID quote is the mid-point between the NPV based on the CDS bid and ask quotes.
More generally, when fitting also more sophisticated hazard rate processes to CDS bid and ask premium quotes we will suffer from the same
problem. The MID feature of the premium rates will not translate in MID NPV's.
Despite this inconvenience, the idea is to use a process for $\lambda$ in the instantaneous spread space to
explain liquidity in the CDS prices.
As a consequence, a formula similar to~(\ref{eq:DEFZCBbis}) is used to price the CDS under liquidity risk.

For example, Chen, Fabozzi and Sverdlove (2008) in the first stage of their investigation fit a constant hazard rate $h$ to $S^{ask}$ and
then calibrate
the liquidity term $\lambda$ (to be subtracted from $h$ when discounting) to reprice the CDS MID quotes. In their view illiquidity always reduces the CDS premium and the CDS premium in the real market is less than a hypothetically perfectly liquid CDS premium. Because the latter is unobservable, they choose as benchmark the ASK CDS premium. The liquidity premium will be reflected in the difference between the transaction CDS premium (MID) and the ASK.
In our example with flat hazard rates and continuous payments in the premium leg, this amounts to set
\[ h^{ask} := \frac{S^{ask}}{\lgd}, \ \ h^{mid} + \lambda =: h^{ask} := \frac{S^{ask}}{\lgd}. \]
This also implicitly suggests that
\[ h^{bid} := \frac{S^{mid}}{\lgd} - \lambda  \] so that again
$\lambda = (h^{ask} - h^{bid})/2$.

CFS also argue that
the liquidity correction should affect only the default leg of the CDS,
since the premium leg is basically an annuity numeraire. This is not clear to us, since in a way also the
premium leg is part of a traded asset subject to mark to market. And indeed,
BT[06,08] make a different choice, and assume that the CDS liquidity discount appears in the payment/fix
leg of the CDS and that there is a bond liquidity discount for the recovery part of the default leg.
We will discuss this more in detail below.

Having clarified the structure of the liquidity term with respect to the usual terms in CDS valuation
with hazard rate/intensity models,
we notice that the above picture considers the CDS bid and ask quotes to be perturbed by liquidity,
so that CDS quotes do not express pure credit risk.

One may also think that the CDS expresses pure credit risk (see for example Longstaff, Mithal and Neis (2005)),
but this is now considered to be unrealistic as illustrated in the papers discussed in this survey.

More generally, we will see that the model for both hazard rates and liquidity are stochastic.
Typically, in the literature, under the risk neutral probability measure,  one assumes stochastic differential equations of diffusion type,
\begin{equation}\label{handlambda} d h_t = b^h(t,h_t) dt + \sigma^h(t,h_t) dW^h_t, \ \ d \lambda_t = b^{\lambda}(t,\lambda_t) dt + \sigma^\lambda(t,\lambda_t) dW^{\lambda}_t,
\end{equation}
where the two processes are possibly correlated. In most works in the literature, though, the two processes are unrealistically
assumed to be independent. 
The advantage of assuming independence is that whenever we need to compute a formula similar to~(\ref{eq:DEFZCBbis}), its relevant component in credit and liquidity can be factored as
\begin{equation}\label{eq:DEFZCBter} \Ex_t
\bigg{[} e^{-\int_t^T ( h_u + \lambda_u ) \ du } \bigg{]}
= \Ex_t \bigg{[} e^{-\int_t^T  h_u \ du } \bigg{]}
\Ex_t \bigg{[} e^{-\int_t^T  \lambda_u  \ du }\bigg{]}
\end{equation}
and this can be computed in closed form whenever we choose processes for which the bond price formulas are known.
Also, in case where no liquidity premium is present, the survival probability $\Qx(\tau > T)$ appearing for example in Formula~(\ref{eq:cdsfairspread0}) is computed as
\begin{equation}\label{eq:survivalhrm} \Qx(\tau > T) = \Ex_0 \bigg{[} e^{-\int_0^T  h_u \ du } \bigg{]}
\end{equation}
and again this is known in closed form, so that CDS fair premium rate can be computed in closed form with the model.

The cases of affine or quadratic models are typical to have the above tractability. However, when we correlate $h$ and $\lambda$, the above decomposition no longer occurs and the only case where calculations are still easy in the correlated case is when the two processes $[h_t,\lambda_t]$ are jointly gaussian. This is not a viable assumption for us, however, since $h$ needs to be positive, being a time-scaled probability.

In order to avoid numerical methods, most authors assume independence between credit risk $(h_t)$ and liquidity risk $(\lambda_t)$ in order to be able to apply formula~(\ref{eq:DEFZCBter}) and the likes.

It is worth highlighting at this point that this assumption is unrealistic, as we will discuss further in Section~\ref{sec:fitchliqscores}. In particular, see Figure~\ref{fig:liqsmile} below.

\subsection{Chen, Fabozzi and Sverdlove (2008) [CFS]}

In this paper, the authors study liquidity and its impact on single name CDS prices for corporations.
From a first exam of the data they notice that the bid-ask spreads are very wide, especially for the high-yield
corporate names in their study. While this is pre-crisis data, they noticed that the liquidity in the CDS market has improved in time, while still maintaining a larger bid-ask spread than typical bid-ask spreads in
the equity market.

After the preliminary analysis, the authors employ a two-factor Cox-Ingersoll-Ross model for the liquidity and hazard rates and estimate their dynamics using Maximum Likelihood Estimation (MLE).

In the above formalism, this means they made two particular choices for the processes~(\ref{handlambda}) to be of the type
\begin{equation}\label{handlambdacir} d h_t = [k^h \theta^h - (k^h + m^h) h_t] dt + \nu^h \sqrt{h_t} dW^h_t, \ \
d \lambda_t = [k^\lambda \theta^\lambda - (k^\lambda + m^\lambda) \lambda_t] dt + \nu^\lambda \sqrt{\lambda_t} dW^{\lambda}_t,
\end{equation}
for positive constants $k^h,\theta^h,\nu^h, h_0, m^h$ and $k^\lambda,\theta^\lambda,\nu^\lambda, \lambda_0, m^\lambda$. The two processes are assumed, somewhat unrealistically (see also the discussion in Section~\ref{sec:fitchliqscores}), to be independent.

The above is the dynamics under the risk neutral or pricing probability measure. This is the dynamics that is used in valuation, to compute risk neutral expectations and prices. The dynamics under the physical measure related to historical estimation and MLE is
\begin{equation}\label{handlambdacirphys} d h_t = k^h (\theta^h -  h_t) dt + \nu^h \sqrt{h_t} d \widetilde{W}^h_t, \ \
d \lambda_t = k^\lambda (\theta^\lambda - \lambda_t) dt + \nu^\lambda \sqrt{\lambda_t} d \widetilde{W}^{\lambda}_t,
\end{equation}
where now $\widetilde{W}$ are brownian motions under the physical measure.
In (\ref{handlambdacirphys}) the $\theta$'s are mean reversion levels, the $k$ are speed of mean reversion parameters, the $\nu$ are instantaneous volatilities parameters. The $m$ parameters are market prices of risk, parameterizing the change of measure from the physical probability measure to the risk neutral one. See for example Brigo and Mercurio (2006). Brigo and Hanzon (1998) hint at the possible application of filtering algorithms and quasi-maximum likelihood to a similar context for interest rate models.

The advantages of the CIR model are that the processes are non-negative, in that $h_t \ge 0$ and $\lambda_t \ge 0$ (and one has strict positivity with some restrictions on the parameters), and there is a closed form formula in terms of $k^h,\theta^h,\nu^h, h_0, m^h$ and $k^\lambda,\theta^\lambda,\nu^\lambda, \lambda_0, m^\lambda$ for
\begin{equation}\label{eq:DEFZCBfour} \Ex_0
\bigg{[} e^{-\int_0^T ( h_u + \lambda_u ) \ du } \bigg{]}
= \Ex_0 \bigg{[} e^{-\int_0^T  h_u \ du } \bigg{]}
\Ex_0 \bigg{[} e^{-\int_0^T  \lambda_u  \ du } \bigg{]}
\end{equation}
and related quantities that are needed to compute the CDS prices.
Indeed, through formula (\ref{eq:cdsfairspread0}) combined with formula
(\ref{eq:survivalhrm}) that is known in closed form for CIR, we have the CDS premium rate in closed form for our model.

Adding the liquidity discount term, $L(t,s) = e^{-\int_t^s \ell_u du}$, into to the CDS premium rate formula (\ref{eq:cdsfairspread}), we obtain

\begin{equation}\label{eq:cdsfairspreadwithliqcfs}
S^{\ast}_{a,b}(t) = \frac{\Ex_t[\lgd\ \mathbf{1}_{\{T_a < \tau \le T_b \} } L(t,\tau) D(t,\tau)]}{\Ex_t[D(t,\tau) (\tau-T_{\beta(\tau)-1})
\mathbf{1}_{\{T_a < \tau < T_b \} }
 +   \sum_{i=a+1}^b D(t,T_i) \alpha_i  \mathbf{1}_{\{\tau \ge T_i\} }]}
\end{equation}

Notice that we added the additional (il)liquidity discount term only in the numerator. This is the strategy followed by CFS, who argue that the annuity should not be adjusted by liquidity. As observed above this is debatable, since even the premium leg of the CDS is part of a traded product, and indeed for example BT[06,08] follow a different strategy.

Formula (\ref{eq:cdsfairspreadwithliqcfs}) can be further explicited in terms of the processes $r, h, \lambda$ via iterated expectations with respect to the sigma field ${\cal F}_{t}$. For the special case of $t=T_a = 0$ one obtains
\begin{eqnarray}\label{eq:cdsfairspreadwithliqcfsexplic}
S^{\ast}_{0,b}(0) = \frac{\Ex_0 \left[ \lgd
\int_{0}^{T_b} h_{u} \exp(-\int_0^u(r_s+h_s+\lambda_{s})ds) du \right]}{ \mbox{Accrual}_{0,b}  +  \sum_{i=a+1}^b \alpha_i \Ex_0 [\exp(-\int_0^{T_i}(r_s+h_s)ds)]   }\\ \nonumber
\mbox{Accrual}_{0,b}  = \int_{0}^{T_b} \Ex_0\left[ h_u \exp\left(-\int_0^u(r_s+h_s)ds\right) (u-T_{\beta(u)-1}) \right]du.
\end{eqnarray}
This holds for general dynamics for $r,h,\lambda$, not necessarily of square root type, and does not require independence assumptions.

In any case, if one sticks to (\ref{eq:cdsfairspreadwithliqcfs}), the CDS fair premium rate formula with deterministic interest rates and hazard rates independent of liquidity spreads reads
\begin{eqnarray}\label{eq:cdsfairspread0liq}
S^{\ast}_{0,b}(0) = \frac{- \lgd
\int_{0}^{T_b} P(0,t)\ A(0,t)\  d_t \Qx(\tau \ge t) }{-\int_{0}^{T_b} P(0,t) (t-T_{\beta(t)-1})
d_t \Qx(\tau \ge t) +   \sum_{i=1}^b P(0,T_i)  \alpha_i   \Qx(\tau \ge
 T_i) }\\ \nonumber
\mbox{where}\\
 A(0,T) = \Ex_{0} [ e^{-\int_0^T \lambda_s ds}] = P^{\mbox \tiny CIR}(0,T; \lambda_0, k^\lambda,\theta^\lambda,\nu^\lambda, m^\lambda)\label{eq:bondlambdacir} , \\
\Qx(\tau \ge T) = \Ex_0 [ e^{-\int_0^T h_s ds}]  = P^{\mbox \tiny CIR}(0,T; h_0, k^h,\theta^h,\nu^h, m^h),
\label{eq:bondhcir}
\end{eqnarray}
where $P^{CIR}$ is the bond price formula in a CIR model having $\lambda$ or $h$ respectively as short rate.
For example,
\begin{eqnarray*}
P^{\mbox \tiny CIR}(0,T; h_0,k^h,\theta^h,\nu^h, m^h) &=& \phi(T-0) \exp(-\psi(T-0) h_0),\\
\phi(T)
  &=& \left[ \frac{2 \sqrt{z} \
       \exp\{({k^h} + {m^h} + \sqrt{z})T/2\} }
      { 2 \sqrt{z}
   + ({k^h} + {m^h} + \sqrt{z}) (\exp\{T \sqrt{z}\} - 1 )}
     \right]^{2 {k^h} \theta^h / (\nu^h)^2} \ , \\
\psi(T)
  &=& \frac{2 (\exp\{T \sqrt{z}\} - 1 )}
  {2 \sqrt{z}
   + ({k^h} + {m^h} + \sqrt{z}) (\exp\{T \sqrt{z}\} - 1 )}\ ,
\\
z &=& ({k^h} + {m^h})^2 + 2 (\nu^h)^2 \ .
\end{eqnarray*}

Notice that we have all the terms in closed form to compute
Formula~(\ref{eq:cdsfairspread0liq}) thanks to the CIR bond price formula and the independence assumption.

Then (\ref{eq:cdsfairspread0liq}) combined with (\ref{eq:bondlambdacir}) and (\ref{eq:bondhcir})   provides a formula for CDS premium rates with liquidity as a function of $h_0,\lambda_0$, and the model parameters $k^h,\theta^h,\nu^h, m^h$ and $k^\lambda,\theta^\lambda,\nu^\lambda, m^\lambda$.
Similarly, formula (\ref{eq:cdsfairspread}) coupled with (\ref{eq:bondhcir}) provided a formula for CDS premium rates without liquidity as a function of $h_0$ and the model parameters $k^h,\theta^h,\nu^h, m^h$.

These formulas can also be applied at a time $t$ later than $0$. Indeed, while taking care of adjusting year fractions and intervals of integration, one applies the same formula at time $t$.

Let us denote the at-the-money liquidity adjusted CDS rate, and the at-the-money CDS rate, at time $t$ by

\[ S^{\ast}_{t,T_b+t}(t;h_t,\lambda_t; k^h,\theta^h,\nu^h, m^h;k^\lambda,\theta^\lambda,\nu^\lambda, m^\lambda), \ \
S_{t,T_b+t}(t;h_t; k^h,\theta^h,\nu^h, m^h),\]
respectively.

A maximum likelihood estimation would then be used ideally, trying to obtain the transition density for
$S^{\ast}_{t+\Delta t,T_b+t+\Delta t}(t+\Delta t;h_{t+\Delta t},\lambda_{t+\Delta t}; k^h,\theta^h,\nu^h, m^h;k^\lambda,\theta^\lambda,\nu^\lambda, m^\lambda)$
given
$S^{\ast}_{t,T_b+t}(t;h_{t},\lambda_{t}; k^h,\theta^h,\nu^h, m^h;k^\lambda,\theta^\lambda,\nu^\lambda, m^\lambda)$
from the non-central (independent) chi-squared transition densities for
$h_{t+\Delta t}$ given $h_{t}$ and $\lambda_{t+\Delta t}$ given $\lambda_{t}$.
One would then maximize the likelihood over the sample period by using such transition densities as key tools, to obtain the estimated model parameters.
Notice that this would be possible only when $h$ and $\lambda$ are independent, since the joint distribution of two correlated CIR processes is not known in closed form.

Similarly, the transition density for
$S_{t+\Delta t,T_b+t+\Delta t}(t+\Delta t;h_{t+\Delta t}; k^h,\theta^h,\nu^h, m^h)$
given
$S_{t,T_b+t}(t;h_{t}; k^h,\theta^h,\nu^h, m^h)$ would be obtained from the chi-squared transition density
for
$h_{t+\Delta t}$ given $h_{t}$.

CFS adopt a  Maximum Likelihood Estimation method based on the earlier work by Chen and Scott (1993). This maximum likelihood method allows CFS  to compute:

\begin{itemize}
\item The credit parameters $k^h,\theta^h,\nu^h, h_0, m^h$ from a time series of ask premium rates
 $S_{0,b}(0)$
\item  The liquidity parameters $k^\lambda,\theta^\lambda,\nu^\lambda, \lambda_0, m^\lambda$ from a time series of mid CDS premium rates $S^{\ast}_{0,b}(0)$.
\end{itemize}

CFS find that the parameters of the hazard rate
factor $h$ are more sensitive to credit ratings and those for the liquidity
component $\lambda$ are more sensitive to market capitalization and  number of quotes, two  proxies for liquidity.

CFS also refer to earlier studies where CDS premiums had been used as a pure measure of the price of credit
risk. CFS argue, through a simulation analysis, that small errors in the CDS credit premium rate
can lead to substantially larger errors in the corporate bond credit spread for the same reference entity. Empirically, they use the CDS estimated hazard rate model above to reprice bonds, with ($h_t$ and $\lambda_t$) and without (just $h_t$) taking CDS liquidity into account.

When using these hazard rates to calculate bond spreads, CFS find that incorporating the CDS liquidity factor
results in improved estimates of the liquidity spreads for the bonds in their sample.

CFS thus argue that while CDS premiums can be used
in the analysis of corporate bond spreads, one must be careful to take into
account the presence of a liquidity effect in the CDS market.

Results reported in the earlier literature before 2006 stated that bond credit spreads were
substantially wider than CDS premiums. This has been contradicted by many observations during the crisis,
but already CFS show that, since a small CDS liquidity premium can translate into a large liquidity discount in a bond's price, mostly due to the principal repayment at final maturity, they can successfully reconcile CDS premiums and bond credit spreads by incorporating liquidity into their model. However, the relevance of this analysis for data in-crisis remains to be proven. Finally, it is worth noticing that in CFS work the (il)liquidity premium is earned by the CDS protection buyer. Indeed, adding a positive (il)liquidity discount rate to the model (and to the default leg only) lowers the fair CDS premium rate with respect to the case with no illiquidity. This means that the protection buyer will pay a lower premium for the same protection in a universe where illiquidity is taken into account, i.e. the liquidity premium is earned by the protection buyer.

\subsection{Buhler and Trapp (2006, 2008) [BT06, BT08]}
BT06 make a different choice, and assume that the CDS liquidity discount appears in the premium leg of the CDS and, furthermore, that there is a bond liquidity discount for the recovery part of the default leg.
BT06 see liquidity as manifesting itself in the bond component of the default leg,
that is involved in the recovery, as one gets a recovery of the bond value ideally when the bond
with face value equal to the CDS notional is meant to be delivered upon default. Their approach may
look debatable as well, although it is further motivated as follows:

``A common solution to this problem both in empirical studies and theoretical models [...],
is to assume that the CDS mid premium is perfectly liquid and thus identical to the transaction premium.
We believe that this assumption, however, is not appropriate. From a theoretical point-of-view,
the assumption suggests that transaction costs, here the bid-ask-spread, are equally divided between
the protection buyer and the protection seller. This fiction neglects the possibility of asymmetric market
frictions which lead to asymmetric transaction costs. The empirical evidence that CDS transaction premia
tend to fluctuate around mid premia, see Buhler and Trapp (2005), adds weight to these theoretical concerns.
In order to reconcile the theoretical arbitrage considerations to a model of CDS illiquidity,
we assume that the mid CDS premium contains an illiquidity component.''

BT06 assume that all the bonds and the CDS for the same issuer have identical default intensity ($h_{t}$) but different liquidity intensities: $\lambda^{b}_{t}$ for all bonds of one issuer and $\lambda^{c}_{t}$ for that issuer CDS. They also assume independence between default free rates, default intensity and liquidity intensities. Using a similar notation as in the previous sections for simplicity, although in their actual work BT06 use discrete time payments in the default leg, the model implied CDS premium rate is equal to
\begin{equation} \label{eq:cdsfairspreadwithliqbt}
S_{0,b}(0) = \frac{-\int_{0}^{T} \ (1-R \ A^{b}(0,t))\ P(0,t) \  d_t \Qx(\tau \ge t) }{-\int_{0}^{T} P(0,t) A^{c}(0,t) (t-T_{\beta(t)-1})
d_t \Qx(\tau \ge t) +   \sum_{i=1}^b  A^{c}(0,T_i) P(0,T_i) \alpha_i   \Qx(\tau \ge
 T_i) }
\end{equation}

 where \[ A^{b}(0,t) = \Ex_{0} [L^{b}(0,t)], \ L^{b}(0,t) = e^{-\int_0^t \lambda^{b}_s ds}, \ \ \ A^{c}(0,t) = \Ex_{0} [ L^{c}(0,t)], \ \ L^{c}(0,t) = e^{-\int_0^t \lambda^{c}_s ds}. \]
 are the liquidity discount factors for the bond and the CDS payment leg respectively.

The default intensity $h_{t}$ is assumed to follow a mean reverting square root process as in (\ref{handlambdacir}). Liquidity intensities are assumed to follow arithmetic Brownian motions with constant drift and diffusion coefficients:
\begin{equation} \label{eq:lintbt}
d \lambda^{j}_t = \mu^{j} dt + \eta^{j} dW^{\lambda^{j}}_t, \ \ j \in \{b,c\}.
\end{equation}

Notation is self-evident. Notice that the (il)liquidity premium will be negative in some scenarios, due to the left tail of the Gaussian distribution for $\lambda$. This is a major difference with CFS, where the illiquidity premium is always positive. These assumptions allow for analytical solutions for bonds and CDS premium rates and for calibration to observed bond spreads and CDS premium rates.

BT06 perform an empirical calibration of the model using bonds and CDS for 10 telecommunications companies over the time period 2001-2005. They find that while the credit risk components in CDS and bond credit spreads are almost identical, the liquidity premia differ significantly. The illiquidity premium component of bond credit spreads is always positive and is positively correlated with the default risk premium. In times of increased default risk bonds become less liquid. The CDS illiquidity premium can take positive or negative values, but is generally very small in absolute value. Contrary to the bonds case, CDS liquidity improves when default risk increases. Thus their framework can explain both positive and negative values for the CDS bond basis through variations in the CDS and bond liquidity. Given the very small sample size in their study, it is not clear whether these results are representative of the whole market. Also, they too use only pre-crisis data. Finally, this approach suffers again from the independence assumption, that is rather unrealistic (see once more the discussion in Section~\ref{sec:fitchliqscores}).

The correlation issue is addressed in BT08, which extends the previous model in BT06 to a reduced form model incorporating now correlation between bond liquidity and CDS liquidity and between default and bond/CDS liquidity. Additionally, they assume different liquidity intensities associated with the ask ($\lambda^{c,ask}_{t}$) and bid CDS ($\lambda^{c,bid}_{t}$).

In the BT08 model the stochastic default intensity ($h_{t}$) and the illiquidity intensities ($\lambda^{b}_{t}$,  $\lambda^{c,ask}_{t}$, $\lambda^{c,bid}_{t}$) are all driven by four independent latent factors $X_{t},Y^{b}_{t},Y^{c,ask}_{t},Y^{c,bid}_{t}$ as follows
    \begin{equation}
    \left(
      \begin{array}{c}
        d h_{t} \\
        d \lambda^{b}_{t} \\
        d \lambda^{c,ask}_{t} \\
        d \lambda^{c,bid}_{t} \\
      \end{array}
    \right) = \left(
                  \begin{array}{cccc}
                    1 & g_{b} & g_{ask} & g_{bid} \\
                    f_{b} & 1 & \omega_{b,ask} & \omega_{b,bid}  \\
                    f_{ask} & \omega_{b,ask} & 1 & \omega_{ask,bid}  \\
                    f_{bid} & \omega_{b,bid} &\omega_{ask,bid} & 1\\
                  \end{array}
                \right)
    \left(
      \begin{array}{c}
        d X_{t} \\
        d Y^{b}_{t} \\
        d Y^{c,ask}_{t} \\
        d Y^{c,bid}_{t} \\
      \end{array}
    \right)
    \label{eq:modelBT08}
\end{equation}

where $X_{t}$ is modeled as a mean reverting square root process as in (\ref{handlambdacir}) and $Y^{b}_{t},Y^{c,ask}_{t},Y^{c,bid}_{t}$ as arithmetic Brownian motions as in (\ref{eq:lintbt}). Again, notation is self-evident. Note that in this model $f$ and $g$ shape the correlations between the default intensity and the liquidity intensities, while $\omega$ shape the liquidity spillover effects between bonds and CDS, which are assumed to be symmetric. Notice that the system of equations in (\ref{eq:modelBT08}) does not guarantee  $L^{c,ask}(0,t) < L^{c,bid}(0,t)$, thus not excluding the case $S_{0,b}^{\mbox \tiny  bid}(0) > S_{0,b}^{\mbox \tiny  ask}(0)$.
However, in their empirical study, they find that this never occurs.

It is further assumed that risk free interest rates are independent of the default and liquidity intensities.

Valuing bonds and CDS in the BT08 framework mainly involves the computation of the expectation of the risk free discount factor and the expectation of the product of the default and liquidity discount factors. The latter expectation is not a product of expectations as before given the assumed dependence between default and liquidity, so that the analogous of Formula~(\ref{eq:DEFZCBfour}) cannot be applied.

The bid CDS premium rate formula becomes:
\begin{equation} \label{eq:cdsfairspreadbidwithliqbt}
S_{0,b}^{\mbox \tiny  bid}(0) = \frac{\int_{0}^{T_b} P(0,t) \Ex_{0} [(1-R e^{-\int_0^t \lambda^{b}_s ds} )  h_t e^{-\int_0^t h_s ds} ] dt }{\sum_{i=1}^{b} P(0,T_i) \alpha_i \Ex_{0} [  e^{-\int_0^{T_i}
\lambda^{c,bid}_s ds}  e^{-\int_0^{T_i} h_s ds} ]+ \mbox{Accrual}  },
\end{equation}
\[\mbox{Accrual} = \int_{0}^{T_b} P(0,t) (t-T_{\beta(t)-1}) \Ex_{0} [ e^{-\int_0^t \lambda^{c,bid}_s ds}   h_t e^{-\int_0^t h_s ds} ] dt \]

The ask CDS premium rate formula is similar with  $\lambda^{c,ask}_t$ instead of $\lambda^{c,bid}_t$. Note that the ask illiquidity discount rate $\lambda^{c,ask}_t$ appears in the payment leg and captures the fact that part of the ask CDS premium rate may not be due to default risk but reflects an additional premium for illiquidity demanded by the protection seller. On the other hand $\lambda^{c,bid}_t$ would capture the illiquidity premium demanded by the protection buyers. Different illiquidity ask and bid spreads reflect asymmetric transaction costs which are driven by the general observed asymmetric market imbalances.

The assumed factor structure of the model and the independence between the latent factors imply an affine term structure model with analytical formulae for both bonds and CDS. For example, expectations in~(\ref{eq:cdsfairspreadbidwithliqbt}) can be computed in closed form.

Data on bonds yields and CDS premium rates on 155 European firms for the time period covering 2001 to 2007 is then used to estimate the model parameters. The estimation procedure generates firm-level estimates for the parameters of the latent variables processes, sensitivities of the different intensities to the latent factors ($f's, g's, w's$) and the values for the credit and liquidity intensities at each point in time ($h_{t}, \lambda^{b}_{t}, \lambda^{c,ask}_{t}, \lambda^{c,bid}_{t}$ ).

The empirical estimation in BT08 implies several interesting findings. First, their results suggest that credit risk has an impact on both bond and CDS liquidity. As credit risk increases, liquidity dries up for bonds and for the CDS ask premium rates ($f_{b}$, $f_{ask}$ are positive and significant). However the impact of increased credit risk on CDS bid liquidity spreads is mixed across different companies, but on average higher credit risk results in lower CDS bid liquidity intensity ($f_{bid}$ is on average negative and significant). Second, their results suggest that while the impact of bond or CDS liquidity on credit risk is negligible ($g_{b}$, $g_{ask}$, $g_{bid}$ are not statistically significant), the spill-over effects between bond and CDS market liquidities are significant ($w_{b,ask}$,$w_{ask,bid}$  are negative and significant, $w_{b,bid}$ is positive and significant). They explain the signs of $w_{b,ask}$,$w_{b,bid}$ as a substitution effect between bonds and CDS: as bond liquidity dries up (bond illiquidity intensity $\lambda^{b}_{t}$ goes up), bond prices go down and thus taking on credit risk using bonds becomes more attractive. If a trader intends to be long credit risk by selling protection through CDS, she will need to drop the ask price (CDS ask liquidity intensity $\lambda^{ask}_{t}$ goes down) compared to the case of high bond liquidity. At the same time lower bond prices in case of lower bond liquidity (higher $\lambda^{b}_{t}$) makes shorting credit risk via bonds more costly which then drives bid quotes in the CDS market higher (higher $\lambda^{bid}_{t}$).

Additionally BT use the empirical parameter and intensity estimates to decompose the bond spreads and CDS premium rates into three separate components: the pure credit risk premium, the pure liquidity risk premium and the rest, the credit-liquidity correlation premium. In particular they estimate the pure CDS credit risk premium ($sd$) as the theoretical CDS premium rate implied by the model when the liquidity intensities $\lambda^{bid}_{t}, \lambda^{ask}_{t}$ are switched off to zero. The pure CDS liquidity premium ($sl$) is subsequently computed as the difference between the average theoretical MID CDS premium rate for uncorrelated credit and liquidity intensities ($f's$ and $g's$ are zero) and the pure CDS credit risk premium $sd$. Finally the correlation premium is calculated as the difference between the observed market CDS premium rate and the sum of the pure credit and pure liquidity premiums.

On average BT08 find that, for CDS, the credit risk component accounts for 95\% of the observed mid premium, liquidity accounts for 4\% and correlation accounts for 1\%. They proceed in similar fashion for the bond spread decomposition and find that overall 60\% of the bond spread is due to credit risk, 35\% is due to liquidity and 5\% to correlation between credit risk and liquidity.

Cross-sectionally all credit, illiquidity and correlation premia for bonds and CDS increase monotonically as the credit rating worsens from AAA to B and then drop for the CCC category. These findings are in contrast to the PTGLKR  findings discussed in Section~\ref{sec:fitchliqscores} and in Figure~\ref{fig:liqsmile} in particular.

BT08 also examine the time series dynamics of the different components. They find that, while generally similar behavior can be observed for the credit risk premium for both investment grade (IG) and high yield (HY) firms, the same is not true for the liquidity premium. During a period with high credit spreads (2001-2002, around Enron and Worldcom defaults) the bond liquidity premium for IG is very volatile and then flattens out at a higher level about mid 2003. On the other hand bond liquidity premium for HY firms reaches the highest level after Worldcom default and decreases to a lower level for the rest of the time period. In the CDS market the CDS liquidity premium for the IG firms is close to 0 for most of time, while for HY it is very volatile and becomes negative when credit risk is high. A negative CDS liquidity premium is consistent with more bid-initiated transactions in the market.

The bond premium dynamics tend to comove over time with the credit risk premium dynamics. Interestingly the correlation premium is larger/smaller than the liquidity premium when credit spreads are high/low. BT interpret this finding as being consistent with the flight to quality/liquidity hypothesis. In other words, in times of stress, investors will try to move away from assets whose liquidity would decrease as credit risk increases and instead acquire liquid assets that can be easily traded. High correlation between illiquidity and credit will thus command a high spread premium component.

All the empirical results with respect to the difference between IG and HY should, in our view, be considered carefully since their sample is highly biased towards investment grade firms. Also, as before, no data in-crisis has been used.

\section{Liquidity through the CAPM framework}

\subsection{Acharya and Pedersen (2005) [AP]}

 There is a fourth work by Bongaerts, De Jong and Driessen (2009) [BDD] who use CAPM like arguments to quantify the impact of liquidity on CDS returns. They construct an asset-pricing model for CDS contracts inspired by the work of Acharya and Pedersen (2005) (AP) that allows for expected liquidity and liquidity risk. Since their approach is heavily based on AP, it is worth recalling AP's general result.

AP start from the fundamental question: ``How does liquidity risk affect asset prices in equilibrium?". This question is answered by proposing an equilibrium asset pricing model with liquidity risk. Their model assumes a dynamic overlapping generations economy where risk averse agents trade securities (equities) whose liquidity changes randomly over time. Agents are assumed to have constant absolute risk aversion utility functions and live for just one period. They trade securities at times $t$ and $t+1$ and derive utility from consumption at time $t+1$. They can buy a security at a price $P_{t}$ but must sell at $P_{t}-C_{t}$ thus incurring a liquidity cost. Liquidity risk in this model is born from the uncertainty about illiquidity costs $C_{t}$. Under further assumptions such as no short selling, AR(1) processes with i.i.d. normal innovations for the dividends and illiquidity costs, AP derive the liquidity-adjusted conditional CAPM:

\begin{eqnarray}\label{eq:LCAPMcond}\nonumber
\underbrace{\mathbb{E}^{P}_{t}\left[R_{t+1}\right]}_{Expected \ Asset \ Gross \ Return} =  \underbrace{R^{f}}_{Risk \ Free \ Rate}+ \underbrace{\mathbb{E}^{P}_{t}\left[c_{t+1}\right]}_{Expected \ Illiquidity \ Cost} \\ \nonumber +\underbrace{\pi_{t}}_{Risk \ Premium} \underbrace{\frac{Cov_{t} \left ( R_{t+1}, R^{M}_{t+1}\right )}{Var_{t} \left ( R^{M}_{t+1}-c^{M}_{t+1} \right )}}_{\beta_{Mkt,t}}
 + \pi_{t}\underbrace{\frac{Cov_{t} \left ( c_{t+1}, c^{M}_{t+1} \right )}{Var_{t} \left ( R^{M}_{t+1}-c^{M}_{t+1} \right )}}_{\beta_{2,t}} \\ \nonumber -\pi_{t}\underbrace{\frac{Cov_{t} \left ( R_{t+1}, c^{M}_{t+1} \right )}{Var_{t} \left ( R^{M}_{t+1}-c^{M}_{t+1} \right )}}_{\beta_{3,t}}-\pi_{t}\underbrace{\frac{Cov_{t} \left ( c_{t+1}, R^{M}_{t+1} \right )}{Var_{t} \left ( R^{M}_{t+1}-c^{M}_{t+1} \right )}}_{\beta_{4,t}}
\end{eqnarray}
where $\pi_{t}=\mathbb{E}^{P}_{t}\left( R^{M}_{t+1}-c^{M}_{t+1}- R^{f}\right)$ is the conditional market risk premium, with the expectation taken under the physical measure. The remaining notation is self-evident.

The liquidity-adjusted conditional CAPM thus implies that the asset's required conditional excess return depends on its conditional expected illiquidity cost and on the conditional covariance of the asset return and asset illiquidity cost with the market return and the market illiquidity cost. The systematic market and liquidity risks are captured by four conditional betas. The first beta ($\beta_{Mkt,t}$) is the traditional CAPM $\beta$ that measures the co-variation of individual security's return with the market return. The second beta ($\beta_{2,t}$) measures the covariance between asset's illiquidity and the market illiquidity. The third beta ($\beta_{3,t}$) measures the covariance between asset's return and the market illiquidity. This term affects negatively the required return. Investors will accept a lower return on securities that have high return in times of high market illiquidity. The fourth beta ($\beta_{4,t}$) measures the covariance between asset's illiquidity and the market return. The effect of this is also negative. Investors will accept a lower return on securities that are liquid in times of market downturns.

In order to estimate the model empirically, the unconditional version of the model is derived under the assumption of constant  conditional covariances between illiquidity and returns innovations. The unconditional liquidity adjusted CAPM can be written as:
\begin{eqnarray}\label{eq:LCAPMuncond}
\mathbb{E}^{P}\left[R_{t}-R^{f}\right]=\mathbb{E}^{P}\left[c_{t}\right]+\pi \beta_{Mkt}+ \pi \beta_{2}-\pi \beta_{3}- \pi \beta_{4}
\end{eqnarray}
where $\pi=\mathbb{E}^{P}[\pi_{t}]$ is the unconditional market risk premium.

AP perform the empirical estimation of the model using daily return and volume data on NYSE and AMEX stocks over the period 1962-1999.  The illiquidity measure for a stock is the monthly average of the daily absolute return to volume ratio proposed by Amihud (2002). Illiquid stocks will have higher ratios as a small volume will have a high impact on price. The Amihud illiquidity measure ratio addresses only one component of liquidity costs, namely the market impact of traded volume. Other components include broker fees, bid-ask spreads and search costs.

Using portfolios sorted along different dimensions, AP find that the liquidity adjusted CAPM performs better than the traditional CAPM in explaining cross-sectional variations in returns, especially for the liquidity sorted portfolios. Liquidity risk and expected liquidity premiums are found to be economically significant. On average the premium for expected liquidity, i.e. the empirical estimate for the unconditional expected illiquidity cost $E(c_{t})$, is equal to 3.5\%. The liquidity risk premium, calculated as $ \pi \beta_{2}-\pi \beta_{3}- \pi \beta_{4}$, is estimated to be 1.1\%. About 80\% of the liquidity risk premium is due to the third component $\pi \beta_{4}$ which is driven by the covariation of individual illiquidity cost with the market return.

\subsection{Bongaerts, De Jong and Driessen (2009) [BDD]}
BDD extend the model proposed by AP to an asset pricing model for both assets in positive net supply (like equities) and derivatives in zero net supply. Differently from the AP framework where short selling is not allowed, in the BDD model some of the agents are exposed to non-traded risk factors and in equilibrium they hold short positions in some assets to hedge these risk factors. Specifically there are two types of assets in the model: basic or non-hedge assets (e.g. equities) which agents hold long positions on in equilibrium and hedge assets which can be held long or short by different agents in equilibrium. Hedge assets are sold short by some agents to hedge their exposures to non-traded risks. Examples of such risks are non-traded bank loans or illiquid corporate bonds held by some financial institutions such as commercial banks. These institutions can hedge the risks with CDS contracts. Other agents such as hedge funds or insurance companies may not have such exposures and may sell CDS to commercial banks to earn the spread.

The BDD model implies that the equilibrium expected returns on the hedge assets can be decomposed in several components: priced exposure to the non-hedge asset returns, hedging demand effects, an expected illiquidity component, liquidity risk premia and hedge transaction costs. Unlike the AP model where higher illiquidity leads to lower prices and higher expected returns, the impact of the liquidity on expected returns in BDD model is more complex. The liquidity risk impact depends on several factors such as heterogeneity in investors' non-traded risk exposure, risk aversion, horizon and agents wealth. Additionally BDD model implies that, for assets in zero net supply like CDSs, sensitivity of individual liquidity to market liquidity ($\beta_{2}$) is not priced.

BDD perform an empirical test of the model on CDS portfolio returns over the 2004-2008 period. The CDS sample captures 46\%
of the corporate bond market in terms of amount issued.
The estimation procedure is a two-step procedure. In the first step, expected CDS returns, liquidity measures (proxied by the bid-ask spread), non traded risk factor returns, non-hedge asset returns and different betas with respect to market returns and market liquidity are estimated. The non-hedge asset returns
are proxied by the S\&P 500 Equity Index Returns. Such estimates represent the explanatory variables of the asset pricing model,
where the response variable is the expected excess return on the hedge asset. In the second step, the generalized method of moments is used to estimate the coefficients of the different explanatory variables in the model.
Their results imply a statistically and economically significant expected liquidity premium priced in the expected CDS returns. On average this is 0.175\% per quarter and it is earned by the protection seller, contrary to CCW and CFS above. They also find that the liquidity risk premium is statistically significant, but economically very small, -0.005\%. Somewhat questionably, the equity and credit risk premia together account for only 0.060\% per quarter.

\section{Predescu et al (2009) [PTGLKR] and Tang and Yan (2007)}\label{sec:fitchliqscores}

Predescu et al (2009) have built a statistical model that associates an ordinal liquidity score with each CDS reference entity.\footnote{Regular commentaries on liquidity scores are available from www.fitchsolutions.com under Pricing \& Valuation Services.}
This provides a comparison of relative liquidity of over 2,400 reference entities in the CDS market globally,
mainly concentrated in North America, Europe, and Asia.
The model estimation and the model generated liquidity scores are based upon the Fitch CDS Pricing Service database, which includes single-name CDS quotes on over 3000 entities, corporates and sovereigns, across about two dozen broker-dealers back to 2000.


The liquidity score is built using well-known liquidity indicators like the bid-ask spread as well as other less accessible predictors of market liquidity such as number of active dealers quoting a reference entity, staleness of quotes of individual dealers and dispersion in mid-quotes across market dealers. The bid-ask spread is essentially an indicator of market breadth; the existence of orders on both sides of the trading book typically corresponds to tighter bid-asks. The other measures are novel measures of liquidity which appear to be significant model predictors for the OTC CDS market. Dispersion of mid quotes across dealers is a measure of price uncertainty about the actual CDS price. Less liquid names are generally associated with more price uncertainty and thus large dispersion.
The third liquidity measure aggregates the number of active dealers and the individual dealers' quote staleness into an (in)activity measure, which is meant to be a proxy for CDS market depth. Illiquidity increases if any of the liquidity predictors increases, keeping everything else constant. Therefore liquid (less liquid) names are associated with smaller(larger) liquidity scores.

\begin{figure}
\includegraphics[width=15cm,height=12cm]{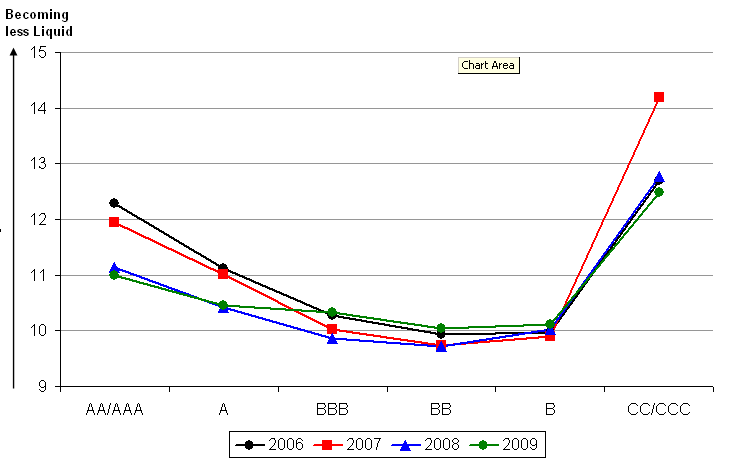}
\caption[liquidity]{Liquidity smile illustrating the correlation between credit quality and liquidity. The vertical axis displays the aggregated values of liquidity scores per rating class. The higher the score the less liquid the name is. } \label{fig:liqsmile}
\end{figure}

The liquidity scores add insight into the liquidity risk of the credit default swap (CDS) market including: understanding the difference between liquidity and credit risk, how market liquidity changes over time, what happens to liquidity in times of stress, and what impact certain credit events have on the liquidity of individual assets. For example it reveals a U-shape relation between liquidity and credit with names in BB and B categories being the most liquid and names at the two ends of credit quality (AAA and CCC/C) being less liquid. (See Figure~\ref{fig:liqsmile}.) The U-shape relation between Liquidity and Credit is quite intuitive as one would expect a larger order imbalance between buy and sell orders for names with a very high or a very low credit rating than for names with ratings in the middle range. In particular, it is reasonable to expect more buying pressure for CCC names and more selling pressure for AAA names.  Most of the trading will take place in the middle categories (BBB, BB, and B). The extent of the illiquidity at the two extremes also changes over time.  This is particularly more pronounced for C-rated entities, which were relatively less liquid in 2007.

Additionally Predescu et al (2009) find that the liquidity score distribution shifts
significantly over the credit/liquidity crisis, having fatter tails (i.e. more names in the very liquid and very illiquid groups) than before the crisis.
The score allows for the construction of liquidity indices at the aggregate market, region or sector levels and, therefore, is very useful for
studying market trends in liquidity.

This study produces an operational measure of liquidity for each CDS reference entity name on a daily basis, and has been extensively validated against external indicators of liquidity.

After the ordinal model by Predescu et al (2009), we go back to works that decompose CDS premium rates levels or changes into liquidity and credit components, and briefly hint at one such study: Tang and Yan (2007). They also include bid ask information among other variables chosen as liquidity measures: volatility to volume, number of contracts outstanding, trade to quote ratio and volume.

They separate liquidity from credit by including other credit control variables in the regression.
The liquidity variables are generally statistically significant, however their impact on premium rates differs.

A more detailed review of Tan and Yang is available in Brigo, Predescu and Capponi (2010).

\section{Discussion, Conclusions and Further Research}
This paper reviews different theoretical and empirical approaches for measuring the impact of liquidity on CDS prices. We start by investigating a reduced form approach that incorporates liquidity as an additional discount yield. The different studies (Chen, Fabozzi and Sverdlove (2008), Buhler and Trapp (2006, 2008)) that use a reduced form model make different assumptions about how liquidity intensity enters the CDS premium rate formula, about the dynamics of liquidity intensity process and about the credit-liquidity correlation. Among these studies BT08 provides the most general and a more realistic reduced form framework by incorporating correlation between liquidity and credit, liquidity spillover effects between bonds and CDS contracts and asymmetric liquidity effects on the Bid and Ask CDS premium rates. However the empirical testing of their model can be significantly improved by using a larger, more representative sample over a longer time period including the crisis.

We then discuss the Bongaerts, De Jong and Driessen (2009) study which derives an equilibrium asset pricing model with liquidity effects. They test the model using CDS data and find that both expected liquidity and liquidity risk have a statistically significant impact on expected CDS returns. However only compensation for expected liquidity is economically significant with higher expected liquidity being associated with higher expected returns for the protection sellers. This finding is contrary to Chen, Cheng and Wu (2005, reviewed in Brigo, Predescu and Capponi, 2010) and Chen, Fabozzi and Sverdlove (2008) that found protection buyers to earn the liquidity premium.

We approach the end of our review with a discussion of Predescu et al (2009) which provides the only operational measure of CDS liquidity that is currently available in the market. They propose a statistical model that associates an ordinal liquidity score with each CDS reference entity and allows one to compare liquidity of over 2400 reference entities. After the ordinal model by Predescu et al (2009), we briefly hint at the work by Tang and Yan (2007), that decomposes again CDS premiums into liquidity and credit components, also including bid ask information among other variables chosen as liquidity measures. This is reviewed more in detail in Brigo, Predescu and Capponi (2010).

Despite their methodological differences, all these studies point to one common conclusion and that is that CDS premium rates should not be assumed to be only pure measures of credit risk. CDS liquidity varies cross-sectionally and over time. More importantly, CDS expected liquidity and liquidity risk premia are priced in CDS expected returns and premium rates. Nevertheless further research is needed to test and validate the CDS liquidity premiums and the separation between credit and liquidity premia at CDS contract level.



\begin{thebibliography}{99}

%
\bibitem{acerbi07}
Acerbi,~C., and Scandolo,~G.~(2008).
Liquidity risk theory and coherent measures of risk.
\emph{Quantitative Finance} 8(7), pp. 681--692.

%
\bibitem{acharya}
Acharya,~V.V., and Pedersen,~L.H.~(2005).
Asset Pricing with Liquidity Risk.
\emph{Journal of Financial Economics} 77(2), pp. 375--410.

%
\bibitem{amihud02}
Amihud,~Y.~(2002).
Illiquidity and Stock Returns: Cross-Section and Time-Series Effects.
\emph{Journal of Financial Markets} 5, pp. 31--56.

%
\bibitem{amihud05}
Amihud,~Y., Mendelson,~H., and Pedersen,~L.H.~(2005).
Illiquidity and Stock Returns: Liquidity and asset pricing.
\emph{Foundation and Trends in Finance} 1, pp. 269--364.

%
\bibitem{angelidis}
Angelidis,~T. and Benos,~A.~(2005).
Liquidity adjusted Value-at-Risk based on the components of the bid-ask spread.
\emph{Working Paper}, available at {\footnotesize \url{http://ssrn.com/abstract=661281}}.

%
\bibitem{bangia}
Bangia,~A., Diebold,~F.X., Schuermann,~T. and Stroughair,~J.D.~(1999).
Modeling Liquidity Risk With Implications for Traditional Market Risk Measurement and Management.
\emph{Working paper}, Financial Institutions Center at The Wharton School.

%
\bibitem{beumee}
Beumee,~J., Brigo,~D., Schiemert,~D., and Stoyle,~G.~(2009).
Charting a Course through the CDS Big Bang.
Fitch Solutions research report, available at {\footnotesize \url{http://www.defaultrisk.com/pp\_crdrv176.htm}}.

%
\bibitem{bielecki01}
Bielecki,~T., and Rutkowski,~M.~(2001).
Credit Risk: Modelling, Valuation and Hedging, Springer, New York, NY.

%
\bibitem{bis08}
BIS~(2008).
Principles for Sound Liquidity Risk Management and Supervision.
Available at {\footnotesize \url{http://www.bis.org/publ/bcbs144.htm}}.

%
\bibitem{bongaerts2009}
Bongaerts,~D., De Jong,~F., and Driessen,~J.~(2009).
Derivative Pricing with Liquidity Risk: Theory and Evidence from the Credit Default Swap market.
Forthcoming in \emph{Journal of Finance}.

%
\bibitem{alfonsi}
Brigo,~D., and Alfonsi,~A.~(2005).
Credit Default Swaps Calibration and Derivatives Pricing with the SSRD Stochastic Intensity Model.
\emph{Finance and Stochastics} 9(1), pp. 29--42.

%
\bibitem{BrigoCapponi}
Brigo,~D., and Capponi,~A.~(2008).
Bilateral counterparty risk valuation with stochastic dynamical models and application to Credit Default Swaps.
Available at {\footnotesize \url{papers.ssrn.com/sol3/papers.cfm?abstract_id=1318024}}.

%
\bibitem{Brigo08}
Brigo,~D., and Chourdakis,~K.~(2009).
Counterparty Risk for Credit Default Swaps: Impact of spread volatility and default correlation.
\emph{International Journal of Theoretical and Applied Finance} 12(7), pp. 1007--1026.

%
\bibitem{BrigoBachir08}
Brigo,~D., and El-Bachir,~N.~(2008).
An exact formula for default swaptions' pricing in the SSRJD stochastic intensity model.
Accepted for publication in \emph{Mathematical Finance}.

%
\bibitem{brigohanzon}
Brigo,~D., and Hanzon,~B.~(1998).
On some filtering problems arising in mathematical finance.
\emph{Insurance: Mathematics and Economics} 22(1), pp. 53--64.

%
\bibitem{brigo2001}
Brigo,~D., and Mercurio,~F.~(2001).
A deterministic-shift extension of analytically tractable and time-homogeneous short rate models.
\emph{Finance and Stochastics} 5(3), pp. 369--388.

%
\bibitem{brigomercurio:2}
Brigo,~D., and Mercurio,~F.~(2001).
Interest Rate Models: Theory and Practice -- with Smile, Inflation and Credit, Second Edition 2006,
Springer Verlag.



\bibitem{brigopredescucapponi} Brigo, D., Presescu, M., and Capponi, A. (2010).
Liquidity modeling for Credit Default Swaps: an overview. To appear in: Bielecki, T., Brigo, D., and Patras, F. (Editors),
 Credit Risk Frontiers: The subprime crisis, Pricing and Hedging, CVA,
MBS, Ratings and Liquidity, Bloomberg Press.



%
\bibitem{brunnpeder}
Brunnermeier,~M., and Pedersen,~L.H.~(2007).
Market Liquidity and Funding Liquidity.
\emph{Review of Financial Studies} 22(6), pp. 2201--2238.


%
\bibitem{buhlertrapp05}
Buhler,~W., and Trapp,~M.~(2005).
A Comparative Analysis of Liquidity in Bonds and CDS markets.
\emph{Working paper}.

%
\bibitem{buhlertrapp06}
Buhler,~W., and Trapp,~M.~(2006).
Credit and liquidity risk in bond and CDS market.
\emph{Working paper}.

%
\bibitem{buhlertrapp08}
Buhler,~W., and Trapp,~M.~(2008).
Time-Varying Credit Risk and Liquidity Premia in Bond and CDS Markets.
\emph{Working paper}.

%
\bibitem{Cetin04}
Cetin,~U., Jarrow,~R., and Protter,~P.~(2004).
Liquidity risk and arbitrage pricing theory.
\emph{Finance and Stochastics} 8(3), pp. 311--341.

%
\bibitem{Cetin04}
Cetin,~U., Jarrow,~R., Protter,~P., and Warachka,~M.~(2005).
Pricing Options in an extended Black-Scholes Economy with Illiquidity: Theory and Empirical Evidence.
\emph{Review of Financial Studies} 19(2), pp. 439--529.

%
\bibitem{ccfl}
Chen,~R.R., Cheng,~X., Fabozzi, F.J.~, and Liu,~B.~(2008).
An Explicit Multi Factor CDS pricing model with correlated factors.
\emph{Journal of Financial and Quantitative Analysis} 43(1), pp. 123--60.

%
\bibitem{ccw}
Chen,~R.R., Cheng,~X., and Wu,~L.~(2005).
Dynamic Interactions Between Interest Rate, Credit, and Liquidity Risks: Theory and Evidence from the Term Structure of Credit Default Swaps.
\emph{Working Paper}, available at {\footnotesize \url{http://papers.ssrn.com/sol3/papers.cfm?abstract_id=779445}}.

%
\bibitem{cfs}
Chen,~R.R, Fabozzi,~F., and Sverdlove,~R.~(2008).
Corporate CDS liquidity and its implications for corporate bond spreads.
\emph{Working paper}, available at {\footnotesize \url{http://business.rutgers.edu/files/cfs20080907.pdf}}.

%
\bibitem{cscott}
Chen,~R.R., and Scott,~L.~(1993).
Maximum Likelihood Estimation of a Multi-Factor Equilibrium Model of the Term Structure of Interest Rates.
\emph{Journal of Fixed Income} 3(3), pp. 14--32.

%
\bibitem{duffie}
Duffie,~D., and Singleton,~K.~(1999).
Modeling Term Structures of Defaultable Bonds.
\emph{Review of Financial Studies} 12(4), pp. 197--226.

%
\bibitem{earnst08}
Earnst,~C., Stange,~S., and Kaserer,~C.~(2009).
Accounting for Non-normality in Liquidity Risk.
Available at {\footnotesize \url{http://ssrn.com/abstract=1316769}}.

%
\bibitem{fsa09}
FSA~(2009).
Strengthening liquidity standards.
Available at {\footnotesize \url{http://www.fsa.gov.uk/pubs/policy/ps09\_16.pdf}}.

%
\bibitem{garleanu09}
Garleanu,~N., Pedersen,~L.H., and Poteshman,~A.M.~(2009).
Demand-based option pricing.
\emph{Review of Financial Studies} 22(10), pp. 4259--4299.

%
\bibitem{jarrow05}
Jarrow,~R., and Protter,~P.~(2005).
Liquidity Risk and Risk Measure Computation.
\emph{Working Paper}, Cornell University.

%
\bibitem{jarrow97}
Jarrow,~R., and Subramanian,~A.~(1997).
Mopping up Liquidity.
\emph{RISK} 10(10), pp. 170--173.

%
\bibitem{johnson}
Johnson,~T.C.~(2008).
Volume, liquidity, and liquidity risk.
\emph{Journal of Finacial Economics} 87(2), pp. 388--417.

%
\bibitem{leippold}
Leippold,~M., and Wu,~L.~(2000).
Quadratic term structure models.
\emph{Working paper}, available at {\footnotesize \url{http://papers.ssrn.com/sol3/papers.cfm?abstract_id=206329}}.

%
\bibitem{longstaff}
Longstaff,~F.A., Mithal,~S. and Neis,~E.~(2005).
Corporate yield spreads: default risk or liquidity? New evidence from the credit default swap market.
\emph{Journal of Finance} 60(5), pp. 2213--2253.

%
\bibitem{morini}
Morini,~M.~(2009).
Solving the puzzle in the interest rate market.
\emph{Working paper}, available at {\footnotesize \url{http://papers.ssrn.com/sol3/papers.cfm?abstract_id=1506046}}.

%
\bibitem{ohara}
O'Hara,~M.~(1995).
Market Microstructure Theory, Blackwell Publishers, Cambridge.

%
\bibitem{Predescu}
Predescu,~M., Thanawalla,~R., Gupton,~G., Liu,~W., Kocagil,~A., and Reyngold,~A.~(2009).
Measuring CDS Liquidity.
Fitch Solutions presentation at the Bowles Symposium, Georgia State University, February 12 2009,
available at {\footnotesize \url{http://www.pfp.gsu.edu/bowles/Bowles2009/Liu-LiquidityMeasure_MethodlogyResults.pdf}}.

%
\bibitem{stange08}
Stange,~S., and Kaserer,~C.~(2008).
Why and How to Integrate Liquidity Risk into a VaR-Framework.
CEFS working paper 2008 No. 10, available at {\footnotesize \url{http://ssrn.com/abstract=1292289}}.

%
\bibitem{szego09}
Szeg\"{o},~G.~(2009).
The Crash Sonata in D Major.
\emph{Journal of Risk Management in Financial Institutions}, 3(1), pp. 31--45.

%
\bibitem{tangyan07}
Tang,~D.Y., and Yan,~H.~(2007).
Liquidity and Credit Default Swap Spreads.
\emph{Working Paper}, available at {\footnotesize \url{http://papers.ssrn.com/sol3/papers.cfm?abstract_id=891263&rec=1&srcabs=676420}}.

\end{thebibliography}
\end{document}